\pdfoutput=1
\documentclass[a4paper,11pt,notitlepage]{article}

\usepackage[margin=20mm]{geometry}
\usepackage{graphicx}
\usepackage{setspace}
\usepackage{amsmath, amssymb, bm}
\usepackage{caption}
\usepackage{siunitx}
\usepackage{multirow}
\usepackage{booktabs}
\usepackage{tabularx}
\usepackage{makecell}
\usepackage{threeparttable}
\usepackage{xspace}
\usepackage{latexsym}
\usepackage[table, dvipsnames]{xcolor}
\usepackage[hidelinks]{hyperref}
\usepackage[capitalise]{cleveref}
\usepackage{soul}
\usepackage{tabularx}
\usepackage{listings}
\usepackage{algorithm}
\usepackage{algorithmic}

\usepackage{times}
\usepackage{helvet}
\usepackage{courier}
\usepackage[most]{tcolorbox}
\tcbset{colback=gray!10,colframe=gray!60!black,boxrule=0.4pt,arc=2pt,
        left=6pt,right=6pt,top=6pt,bottom=6pt}
\newtcolorbox{codebookbox}[1][]{breakable,#1}

\newcommand{\keywords}[1]{\textbf{\textit{Keywords:}} #1}

\usepackage{xcolor}

\begin{document}
 \begin{center}
   \LARGE{\textbf{Measuring political fragmentation in networks}}
 
\vspace{1cm}

\large Yuan Zhang$^{1,\ast}$, Laia Castro$^{2}$, Frank Esser$^{1}$ and Alexandre Bovet$^{3,4}$

\vspace{0.2cm} \normalsize 
$^{1}$Department of Communication and Media Research, University of Zurich, Zurich, Switzerland\\
$^{2}$Department of Political Science, University of Barcelona, Barcelona, Spain\\
$^{3}$Department of Mathematical Modeling and Machine Learning, University of Zurich, Zurich, Switzerland\\
$^{4}$Digital Society Initiative, University of Zurich, Zurich, Switzerland\
$\ast$ y.zhang1227@hotmail.com

\end{center}

\abstract{Understanding the emergence of fragmentation in complex systems, such as social and political networks, is an interdisciplinary challenge requiring the precise measurement and comparison of fragmentation across empirical contexts. Here, we show that common metrics of network fragmentation, such as modularity and the Krackhardt E--I index, yield inconsistent results as they conflate two aspects of fragmentation: the strength of separation and the number of fragments. These two aspects have not been clearly distinguished in previous studies, making comparisons across varied political systems difficult to interpret. In addition, none of them is designed to capture the multiscale fragmentation structures that characterize real-world complex social systems. We compare several widely used network measures and show that the two aspects of network fragmentation are best captured by the pairwise adaptive E--I index and the effective number of communities. Furthermore, we introduce a novel metric for multiscale fragmentation, the effective branching factor, which captures the degree to which communities redistribute across finer levels of network partition and applies to hierarchical and non-hierarchical structures. We demonstrate the application of the three measures in two cross-national online political interaction networks in Brazil, Spain, and the United States. Overall, these results clarify three complementary dimensions of network fragmentation: strength of separation, number of fragments, and between-level branching. While our application is limited to the case of political interaction networks, our approach has the potential to be applicable to complex networks representing other systems.}


\section{Introduction}
\label{sec1}

Network fragmentation refers to the differentiation of a network into multiple
groups and the extent of separation among them, and is associated with
the concept of network modularity \cite{newman2006modularity}. It is a
common structural feature of many complex networks, including social
\cite{pham2020effect}, political
\cite{bakshy2015exposure,barbera2015tweeting,cinelli2021echo}, biological
\cite{alcala2021modularity}, technological and information systems
\cite{xue2013risk}. In online social networks, for instance, fragmentation
is associated with the existence of echo chambers of political opinions.
When opinions become extreme or antagonistic to other groups, it can undermine
trust in institutions, erode democratic norms, and impede governance by
making political cooperation more difficult
\cite{cinelli2020selective,mccarty2016polarized,iyengar2019origins}.
How online fragmentation develops is a question that continues to challenge
contemporary research.

Despite its importance, network fragmentation is difficult to measure and
compare across networks. A network may be fragmented because it has more
cohesive communities or because it has a larger number of fragments. Current
measures usually confound these dimensions. Existing network measures include
modularity \cite{conover2011political} and the Krackhardt E--I index and
its variants
\cite{krackhardt1988informal,bright2018fragmentation,chen2024climate}.
Modularity compares the observed density of within-community edges with
the expected density under a randomized null model
\cite{newman2004finding}. However, modularity-based approaches can be sensitive
to network and community sizes and are subject to a well-known resolution
limit: when a network is sufficiently large, small yet meaningful communities
may fail to be detected \cite{fortunato2010community}. The Krackhardt E--I
index (we refer to it as the global E-I index later) captures the relative
balance of the between-group (external) ties and within-group (internal)
ties in a clustered network \cite{krackhardt1988informal}. It has also
been widely used to calculate the connection strength of a single community
within a network, where between-group ties are defined as ties between
that community and all other communities (we refer to it as the community-level
E-I index later)
\cite{nuernbergk2016conversations,chin2022evaluating,bruns2017echo}.
However, the E--I index is also sensitive to both network and community
sizes \cite{bright2018fragmentation}. To mitigate the sensitivity, the
E--I index has been extended from its global- and community-level formulations
to pairwise and propensity-based variants \cite{chen2021polarization}.

Another approach estimates the effective number of communities (ENC) by
accounting for size inequalities within a given partition. The metric originates
in political science, where it was introduced to measure the effective
number of parties based on electoral vote shares or seat allocation
\cite{laakso1979effective,golosov2010effective} and is also used in ecology
to measure species diversity \cite{chao2016phylogenetic}. As community
detection algorithms have enabled the automatic identification of communities,
ENC has also been adopted in network analysis
\cite{onnela2012taxonomies}. Unlike modularity and the E-I index, ENC aims
to capture the number of fragments rather than the strength of their separation.

Beyond conceptual conflicts in existing fragmentation metrics, neither
effective number measures nor network-based measures of separation strength
capture the multiscale structure of network fragmentation. Multiscale structure
refers to the organization of a network into distinct community partitions
across multiple resolutions, such that different real-world patterns of
social and political grouping may become salient at different levels rather
than being fully represented by a single level
\cite{delvenne2010stability}. Many existing related measures focus on hierarchical
structures, which can be regarded as a special case of multiscale organization
but require a stricter assumption that communities are nested across successive
levels \cite{schaub2023hierarchical}. A major advance in multiscale networks
is multiscale community detection, which enables identifying clusters across
multiple resolution levels
\cite{arenas2008analysis,peixoto2014hierarchical,delvenne2010stability,arnaudon2024pygenstability}.
Yet none of the metrics discussed above can measure multiscale fragmentation
in a way that is comparable across contexts. To address this gap, we introduce
a novel fragmentation metric, the \emph{effective branching factor} (EBF),
that integrates the effective number of communities with the branching
factor to measure between-level branches in a multiscale political space.
Drawing on social interpretations, the metric quantifies the extent to
which one dimension of division combines with other dimensions to produce
finer-grained clusters \cite{zhang2025multilevel}.

Using the planted partition model (PPM), we demonstrate that modularity,
the global-level E--I index, and the community-level E--I index are influenced
by both the number of communities and the relative ratio of within- to
between-community edge probabilities. We also show, using stochastic block
model (SBM) simulations, that the pairwise E-I index is affected by imbalances
in community sizes. Our results suggest that only using the pairwise adaptive
E--I index and the ENC can independently capture the separation strength
and the effective number of groups in the single-scale fragmentation. Moreover,
only our EBF metric can quantify structural political fragmentation in
the multiscale setting.

We then apply the three measures (pairwise adaptive E--I index, ENC, and
EBF) to two empirical datasets in political contexts across countries:
(1) ordinary users' political news-consumption traces and (2) ordinary
users' Twitter/X following networks of political influencers. Both datasets
were collected during election periods in Brazil (2022), Spain (2023),
and the United States (2024) using the same research design. Applying EBF
to the two empirical datasets, we find that the average EBF values yield
similar cross-national comparisons of structural political fragmentation
across Brazil, Spain, and the United States -- Brazil shows the highest
level of structural political fragmentation, the United States the lowest,
with Spain in the middle. However, examining the number of fragments (measured
by ENC) and the separation strength (measured by the pairwise adaptive
E-I index) at each level reveals differences between the two datasets.

This study contributes to the development of quantitative metrics for measuring
network fragmentation and structural political fragmentation. Although
many metrics have been used for this purpose, they in fact capture different
dimensions of fragmentation, such as separation strength and the number
of fragments, and no single metric can represent all of them. Our work,
therefore, provides a theoretical interpretation of the dimensions of network
fragmentation captured by existing measures and proposes a new approach
that can also quantify between-level branching under multiscale conditions.

\subsection{Political fragmentation in the digital age}
\label{sec1.1}

In recent decades, online political environments have undergone a significant
transformation and therefore provide an appropriate test case for network
fragmentation measures.

Long before the emergence of the Internet, Blau's theory of multidimensional
social structure conceptualized society as organized simultaneously along
multiple intersecting social dimensions; the rise of digital communication
has subsequently amplified the visibility, interaction, and combination
of these dimensions in everyday social life
\cite{blau1977macrosociological}. The twentieth century witnessed an explosion
of information due to the development of the World Wide Web. The Internet
has weakened traditional geographic constraints on information dissemination
and intensified competition for attention by enabling increasingly targeted
content \cite{pariser2011filter}. Van Alstyne and Brynjolfsson
\cite{van1996electronic} describe this process as multi-dimensional ``stratification,''
in which individuals are clustered not only by geography but also by additional
dimensions, such as topic interests. This transformation is also reflected
in the political domain \cite{cinelli2021echo}. The digital world has revealed
political fragmentation that predated the Internet \cite{poole1984polarization}. For instance, online news consumers tend to spend more time on news sources
that align with their own political views, consistent with polarization along the left--right dimension \cite{garimella2021political}.

One of the most significant changes of the digital age is that individual
users have gained greater power in the dissemination of information
\cite{castells2007communication}. For instance, since the rise of Donald
Trump, an increasing number of politicians have relied on personalized,
candidate-centered messaging during election campaigns
\cite{papathanassopoulos2025political}. Similarly, an increasing number
of journalists can now build independent channels and express opinions
with fewer institutional constraints \cite{kedem2024journalists}. Moreover,
a large number of influencers have emerged as meso-level actors (neither
official institutions nor ordinary users) who circulate and interpret political
information \cite{goodwin2023political}. Across these individual channels,
greater audience attention increases the likelihood of electoral success
and the potential for monetary gain
\cite{wang2024can,goodwin2023political}. Additionally, platform algorithms
further facilitate personalized recommendations from producers to targeted
audiences, a dynamic commonly described as the ``filter bubble''
\cite{pariser2011filter}. In the competition for audience attention among
an increasing number of political information providers and platforms,
increasingly idiosyncratic topics are used to attract audiences. Research
has shown that social identities (e.g., gender, race, religion) and specific
issues (e.g., abortion, gun control, environmental protection) are frequently
invoked in electoral messaging
\cite{holman2015gender,schwor2024religious,sinno2022political,conover2011political}.

People who used to connect primarily around a single shared topic can now
connect with others across many different topics. Each person can be seen
as having a set of political considerations \cite[pp.~40--52]{zaller1992nature}. People who share similar interests are more likely to interact, and different combinations of shared interests can give rise to different patterns of connections. Interaction-based political communities can form at different levels based on these connection patterns. Depending on which
dimensions of shared interests underlie community formation, some partition levels
may consist of smaller, more granular communities, whereas others may comprise
broader communities. Previous work showed that these partition levels are
linked to the multiscale structure of online networks
\cite{zhang2025multilevel}.

\subsection{Network measures of structural political fragmentation}
\label{sec1.2}

Digital platforms generate increasingly diverse forms of interaction data
from political users, which has been studied as structural political fragmentation
by both network science and social sciences domains
\cite{bright2018fragmentation,salloum2022separating,bruns2023public}.
A wide range of network metrics has been used to quantify structural political
fragmentation. However, most of these measures were developed to capture
polarization, understood as antagonism between two clusters, whereas relatively
few address fragmentation, which may involve multiple clusters and scales.

Community detection algorithms are commonly employed to identify clusters
within network data, where users interact more frequently with others in
the same cluster than with those in different clusters
\cite{fortunato2010community}. Once clusters (communities) are found, network
metrics are used to characterize their separation strength. Modularity
and the Krackhardt E--I index are among the most common metrics for this
purpose. Modularity captures the extent to which a network is divided into
groups with denser internal ties and sparser external ties than expected
under a null model \cite{newman2006modularity}. Because its values depend
on features such as community quantity and community size, researchers
often use it for comparison with additional precautions. For example, Conover
et al. \cite{conover2011political} evaluated modularity in U.S. political
Twitter by comparing observed scores with null-model baselines, while Maj\'{o}-V\'{a}zquez
et al. \cite{majo2019backbone} used modularity on backbone-extracted audience
networks to compare fragmentation across countries. Yet these strategies
do not address the underlying limitations of the metrics.

The Krackhardt E--I index provides a related way to assess fragmentation
by comparing external ties with internal ties
\cite{krackhardt1988informal}. It has been used both for entire networks
and for individual communities in political communication research
\cite{van2021political,nuernbergk2016conversations,chin2022evaluating,bruns2017echo}.
However, this measure is also shaped by network structure, especially the
number of communities and differences in group size. In response, later
studies proposed modified versions and introduced them to the political
domain. Bright \cite{bright2018fragmentation}, for instance, employed a
pairwise fragmentation score for party pairs and showed that it was still
sensitive to imbalances in activity between groups. Chen et al.
\cite{chen2021polarization} introduced the pairwise adaptive E--I index, which uses
within-group and between-group densities rather than raw tie counts to
reduce bias from unequal group sizes. Even so, Salloum et al.
\cite{salloum2022separating} show that such adjustments do not remove all
structural bias, and argue that denoising or normalization against null
models remains necessary.

Additionally, the ENC is a simple measure adapted from the effective number
of parties that captures only the normalized number of communities in a
network. Researchers have also applied it to measure political fragmentation,
but it reflects a different aspect of fragmentation than modularity and
the E-I index family \cite{brito2020complex}.

Moreover, as discussed above, online political environments often exhibit
multiscale structures, in which individuals belong to different clusters
at different levels of connectivity
\cite{rosvall2011multilevel,fortunato2016community}. While several works
focus on characterizing hierarchical networks with tree-balance indices
\cite{lemant2022robust,boucherie2025adaptive}, they do not capture their
level of fragmentation. Moreover, strict hierarchical organization is a
special case of multiscale organization: each finer community belongs entirely
to a single coarser community. Multiscale analysis examines community structure
across resolution levels without requiring this nesting constraint
\cite{lambiotte2015random}. Consequently, nested subdivisions in some parts
of a network can coexist with changes in community composition elsewhere.
This broader perspective accommodates structures that cannot be represented
by a single dendrogram. Within this setting, our work aims to quantify
how the members of each community are distributed across communities at
the next finer resolution, capturing both nested subdivision and non-nested
reconfiguration. Recent advances in community detection enable analyses
across multiple resolutions, capturing political groupings of varying sizes.
For example, hierarchical stochastic block models extend the stochastic
block model framework for the statistical inference of nested structure
\cite{peixoto2014hierarchical}, and random-walk-based approaches enable
the detection of communities at multiple scales via diffusion processes
on networks
\cite{delvenne2010stability,rosvall2011multilevel,schaub2012encoding,arnaudon2024pygenstability}.
However, metrics for quantifying and comparing structural political fragmentation
in multiscale community structures remain underdeveloped.

Building on previous work, we contribute to research on political network
fragmentation by comparing measures of the separation strength and the
number of fragments. We also propose to use the effective branching factor
(EBF) to accommodate multiscale structures without strict nesting. Using
empirical political networks, we demonstrate how these complementary perspectives
reveal distinct patterns of fragmentation.

\section{Theoretical framework}
\label{sec2}

\subsection{Comparison of network fragmentation measures in the single-scale case}
\label{sec2.1}

In the single-scale setting, we compare six commonly used structural fragmentation
metrics: modularity \cite{conover2011political,majo2019backbone}, the
global-level E--I index \cite{van2021political}, the community-level E--I
index
\cite{nuernbergk2016conversations,chin2022evaluating,bruns2017echo},
the pairwise E--I index \cite{bright2018fragmentation}, the pairwise adaptive
E--I index \cite{salloum2022separating}, and ENC
\cite{brito2020complex}. We define them below.

Given an undirected network $G(\mathcal{V},\mathcal{E})$ with node set
$\mathcal{V}$ comprising $N$ nodes, an edge set $\mathcal{E}$, and a partition
of the node set into $L$ non-overlapping communities, modularity quantifies
how much more densely connected nodes are within communities than expected
under a degree-preserving null model
\cite{newman2004finding,newman2006modularity} and is defined as
\begin{equation*}
Q \;=\; \frac{1}{2m}\sum _{i,j}\left (A_{ij}-\frac{k_{i} k_{j}}{2m}
\right )\mathbf{1}(g_{i}=g_{j}).
\end{equation*}

Here, $A_{ij}$ is the adjacency matrix entry indicating whether there is
an edge between nodes $i$ and $j$ (or the weight of that edge in a weighted
network), $k_{i}$ and $k_{j}$ are the degrees of nodes $i$ and $j$,
$m$ is the total number of edges in the network, $g_{i}$ and $g_{j}$ denote
the community memberships of nodes $i$ and $j$, and
$\mathbf{1}(g_{i}=g_{j})$ is an indicator function equal to 1 when nodes
$i$ and $j$ belong to the same community and 0 otherwise.

The global-level E--I index summarizes the overall separation between communities
by contrasting between- vs within-community edges
\cite{krackhardt1988informal}. It is defined as
%
\begin{equation}
\mathrm{EI}_{\text{global}} \;=\; \frac{E-I}{E+I},
\label{eq1}
\end{equation}
where $I=\sum _{(i,j)\in \mathcal{E}}\mathbf{1}(g_{i}=g_{j})$ and
$E=\sum _{(i,j)\in \mathcal{E}}\mathbf{1}(g_{i}\neq g_{j})$ denotes the
total number of internal (within-community) edges and external (between-community)
edges.

The community-level E--I index measures the separation for each community
by contrasting ties that leave the community (to all other communities)
with ties within the community \cite{hargittai2008cross}. For community
$r \in \{1,\dots ,L\}$, define
$ I_{r}=\sum _{(i,j)\in \mathcal{E}}\mathbf{1}(g_{i}=r,\, g_{j}=r)$ and
$E_{r}=\sum _{(i,j)\in \mathcal{E}}\left [\mathbf{1}\left (g_{i}=r,\, g_{j}
\neq r\right )+\mathbf{1}\left (g_{i}\neq r,\, g_{j} = r\right )
\right ]$. Its community-level E--I index is
%
\begin{equation}
\mathrm{EI}_{\text{community}}(r)\;=\;\frac{E_{r}-I_{r}}{E_{r}+I_{r}}.
\label{eq2}
\end{equation}
Although previous studies have primarily reported the community-level E-I
index separately for each community, this measure can also be extended
to the global level by averaging it across all communities.

The pairwise E--I index quantifies the separation between pairs of communities
in a network \cite{bright2018fragmentation}. For communities
$r\neq s$, let
$I_{r}=\sum _{(i,j)\in \mathcal{E}}\mathbf{1}(g_{i}=r,\, g_{j}=r)$ and
$E_{rs}=\sum_{(i,j)\in\mathcal{E}}
\bigl[\mathbf{1}(g_i=r,\,g_j=s)+\mathbf{1}(g_i=s,\,g_j=r)\bigr]$.

The pairwise E--I index is
%
\begin{equation}
\mathrm{EI}_{\text{pair}}(r,s)\;=\;
\frac{E_{rs}-(I_{r}+I_{s})}{E_{rs}+(I_{r}+I_{s})}.
\label{eq3}
\end{equation}
To quantify pairwise fragmentation across the entire network, the average
over all pairs of communities can be employed.

The adaptive (density-based) E--I index is computed by comparing link densities
(or tie propensities) rather than raw counts to mitigate the effect of
communities of varying sizes on E--I indices
\cite{salloum2022separating,chen2021polarization}. This index has been
used to measure structural polarization, i.e., when the network is split
into only two communities, but any of the E--I versions introduced above
can be adapted using densities. We consider here the pairwise adaptive
E--I index, computed for two communities $r$ and $s$, as
%
\begin{equation}
\mathrm{EI}_{\text{adapt}}(r,s) =
\frac{d_{rs}+d_{sr}-(d_{rr}+d_{ss})}{d_{rs}+d_{sr} + d_{rr}+d_{ss}},
\label{eq4}
\end{equation}
where $d_{rr} = I_{r} \binom{n_{r}}{2}^{-1}$ is the within-community density
of community $r$, and $d_{rs} = d_{sr} = E_{rs}(n_{r} n_{s})^{-1}$ is the
link density between communities $r$ and $s$ of sizes $n_{r}$ and
$n_{s}$, respectively. To stay consistent with the previous E--I indices,
we use the opposite form, external density minus internal density, of the
original adaptive E--I index from
\cite{chen2021polarization,salloum2022separating}.

The effective number of communities (ENC) quantifies the number of communities
that are effectively present, accounting for imbalances in community sizes.
Let $p_{r} = n_{r}/N$ be the proportion of nodes belonging to community
$r$. Then the ENC is
%
\begin{equation}
\mathrm{ENC} = \frac{1}{\sum _{r=1}^{L} p_{r}^{2}}
\label{eq:enc}
\end{equation}

It equals $L$ when all communities are equal in size and decreases when
sizes are unequal. This measure is also known as the inverse Simpson index
\cite{simpson1949measurement}.

To understand which aspects of fragmentation these measures capture, we
derive their expressions in the simplest network model of fragmentation
using the planted partition model (PPM) \cite{condon2001algorithms}. This
model is a special case of the stochastic block model where the network
is split into $L$ equal-sized communities and the probabilities of connection
for two nodes in the same community, $p_{\mathrm{in}}$, or in different
communities, $p_{\mathrm{out}}<p_{\mathrm{in}}$, are uniform.

\begin{table*}
\caption{Expressions of the six fragmentation metrics in the planted partition
model (PPM) in the limit of large communities ($N/L\gg 1$). We use
$\rho =p_{\mathrm{in}}/p_{\mathrm{out}}$ for the in-to-out edge probability ratio}
\label{tab:ppm_metrics_summary}
\begin{tabular*}{\textwidth}{@{\extracolsep{\fill}}l l l l@{}}
\hline
Metric & PPM ($\rho =p_{\mathrm{in}}/p_{\mathrm{out}}$) & Relation with $L$ & Relation with $\rho $ \\
\hline
Modularity             & $\displaystyle \frac{L-1}{L}\, \frac{\rho -1} {\rho +L-1} $           & concave with maximum at $L^{\star}=1+\sqrt{\rho}$ & concave increasing \\
[9pt]Global-level E--I      & $\displaystyle \frac{L-1-\rho} {L-1+\rho} $                           & concave increasing                                & convex decreasing  \\
[9pt]Community-level E--I   & $\displaystyle \frac{(L-1)-\frac{1}{2}\rho} {(L-1)+\frac{1}{2}\rho} $ & concave increasing                                & convex decreasing  \\
[9pt]Pairwise E--I          & $\displaystyle \frac{1-\rho} {1+\rho} $                               & independent                                       & convex decreasing  \\
[9pt]Pairwise adaptive E--I & $\displaystyle \frac{1 - \rho}{1 + \rho} $                            & independent                                       & convex decreasing  \\
[9pt]ENC                    & $L$                                                                   & linear increasing                                 & independent        \\
 \hline
\end{tabular*}
\end{table*}

We show the expressions of each network measure in Tab.~\ref{tab:ppm_metrics_summary}
in the limit of large communities ($N/L\gg 1$) for the PPM. The results
are also visualized in Fig.~\ref{fig:comparison_plot}a \& b. The details
of the derivations are shown in Sect.~1 of the Supplementary Information.
The results reveal that modularity, global- and community-level E--I all
depend both on L and $\rho =p_{\textrm{in}}/p_{\textrm{out}}$. This makes
them unsuitable for measuring any single aspect of fragmentation. In particular,
we show that Modularity is ill-suited for comparing the fragmentation of
different networks. It is a concave function of $L$ which reaches a maximum
value at $L=1 + \sqrt{\rho}$. This indicates that, for example, two networks
with the same values of $p_{\textrm{in}}$ and $p_{\textrm{out}}$ but different
numbers of communities can have very similar modularity values. On the
other hand, we show that, in the large-community limit, the pairwise E--I
index is independent of $L$ and it is equal to the pairwise adaptive E--I
index in the PPM. In the PPM, the pairwise (or pairwise adaptive) E--I index and
the ENC therefore provide two independent complementary measures of network
fragmentation. The pairwise (or pairwise adaptive) E--I index measures separation
strength while the ENC measures the number of fragments.

In addition, we examine how heterogeneity in community sizes affects the
six metrics through stochastic block model simulations in which community
sizes are drawn from a power-law distribution (see the simulations in Sect.~2 of the Supplementary Information). Fig.~\ref{fig:comparison_plot}c
\& d show how the ENC and the pairwise and pairwise adaptive E-I indices
vary as a function of the community size heterogeneity. Under unequal-sized
conditions, the pairwise E--I index is no longer independent of community
sizes.

\begin{figure}[htbp]
\centering
\includegraphics[width=\textwidth]{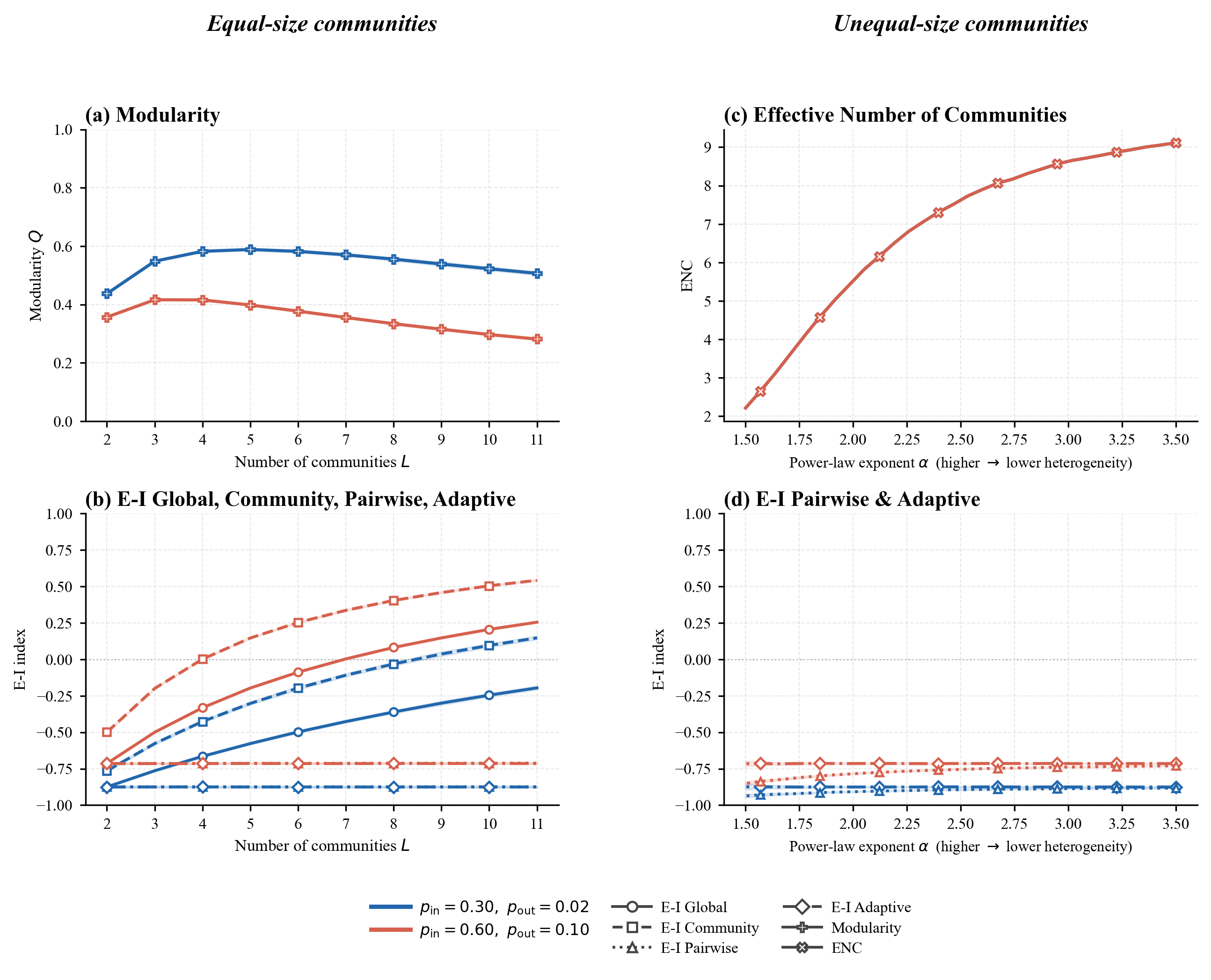}
\caption{Single-scale fragmentation metrics under two SBM settings ($p_{
\mathrm{in}}=0.30$, $p_{\mathrm{out}}=0.02$, blue; $p_{\mathrm{in}}=0.60$,
$p_{\mathrm{out}}=0.10$, red; $N=1{,}000$). Left: equal-sized communities varying
in $L$; (a) modularity $Q$, (b) four E-I variants. Right: power-law community
sizes ($L=10$) varying in $\alpha $; (c) ENC, (d) E-I Pairwise and Adaptive}
\label{fig:comparison_plot}
\end{figure}

In summary, in the PPM, modularity, the global-level E--I index ($
\mathrm{EI}$), and the community-level E--I index ($\mathrm{EI}_{c}$) are
all influenced by the number of communities ($L$) (see Fig.~\ref{fig:comparison_plot} a \& b). More specifically, modularity first
increases with the number of communities and then decreases after reaching
a maximum; the global- and community-level E--I indices increase with the
number of communities. In contrast, the pairwise E--I index (under the
large-$n$ approximation) and the pairwise adaptive E--I index depend only on the
relative strength of within-community connections ($p_{\mathrm{in}}/p_{
\mathrm{out}}$) and therefore can be used to measure separation strength.
The pairwise adaptive E--I index is preferable to the pairwise E--I index
because it yields comparable values even when communities differ in size,
whereas the pairwise E--I index is affected by those size differences (see
Fig.~\ref{fig:comparison_plot} d). On the other hand, the ENC is independent
of the separation strength and provides a flexible measure of the number
of fragments, which is also adapted to unequal community sizes (see the
simulations in Sect.~2 of the Supplementary Information).

Our results show that different metrics capture different dimensions of
structural fragmentation, and no single index covers all of them. We show
that, under a simplified stochastic block model, the pairwise adaptive
E--I and the ENC provide two independent measures of connection strength
and number of fragments, while the other measures confound these two aspects
of fragmentation. However, these metrics are defined for a single-scale
network partition and therefore cannot capture cross-scale relationships
when multiple scales are present. Next, we introduce how our approach (EBF)
captures across-scale fragmentation.

\subsection{Development of a multiscale network fragmentation measure}
\label{sec2.2}

To capture multiscale network fragmentation, we introduce a metric that
integrates both intra-level diversity and inter-level branching. This is
because fragmentation in multiscale structures is not fully described by
diversity at any single level, but also depends on how divisions branch
and propagate across levels. Intra-level diversity quantifies heterogeneity
within a given level, whereas inter-level branching captures how fragmentation
expands or consolidates across levels. Integrating these two components,
therefore, yields a measure of between-level branching capacity.

Note that we have used the terms scale and level interchangeably so far;
we now clarify that the scale is a continuous parameter that can be used
to scan from fine resolutions to large resolutions in a network, while
the levels are discrete and correspond to specific scale values selected
by a multiscale community detection method as the set of scales that best
capture the multiscale nature of the network. Let
$\ell = 1, \ldots , \Gamma $ denote the levels of the network's multiscale
community partition, where $\ell =1$ is the coarsest level and
$\ell =\Gamma $ is the finest. At a given level~$ \ell $, let ~$\mathcal{C}^{(\ell )} = \{C_{1}^{(\ell )}, C_{2}^{(
\ell )}, \ldots , C_{n_{\ell}}^{(\ell )}\}$ denote the partition of the nodes into $n_\ell$ communities. The fragmentation within this
level can be measured using the effective number of communities, ENC, (see
Eq.~(\ref{eq:enc})). We do not require the partitions to form a strict hierarchy;
however, we assume that partitions are sorted from the one with the smallest
effective number of communities (ENC) ($\ell =1$) to the one with the largest
effective number of communities ($\ell =\Gamma $).

To account for multi-level fragmentation, we use the ENC to define an effective
branching factor, which quantifies how communities subdivide into finer-grained
sub-communities. Note that even if branching factors are usually defined
for trees, i.e., for hierarchical structures, our definition is also valid
for non-hierarchical structures.

For a community $C^{(\ell )}_{i}$, its effective branching factor
$B^{(\ell )}_{i}$ from level $\ell $ to level $\ell +1$ is given by the
effective number of communities in which it is subdivided:
\begin{equation*}
B_{i}^{(\ell )} =\left ( \sum _{j=1}^{n_{\ell +1}} \left (
\frac{|C^{(\ell )}_{i} \cap C^{(\ell +1)}_{j}|}{|C^{(\ell )}_{i} |}
\right )^{2}\right )^{-1}.
\end{equation*}

We can then measure the fragmentation from level $\ell $ to
$\ell +1$ as the weighted average of the effective branching factors of
the communities from $\ell $ to $\ell +1$:
%
\begin{equation}
\text{FRAG}(\ell \rightarrow \ell +1) = \sum _{i=1}^{n_{\ell}}
\frac{|C^{(\ell )}_{i} |}{N} \cdot \left ( \sum _{j=1}^{n_{\ell +1}}
\left (
\frac{|C^{(\ell )}_{i} \cap C^{(\ell +1)}_{j}|}{|C^{(\ell )}_{i} |}
\right )^{2}\right )^{-1},
\label{eq:frag_l}
\end{equation}
where $N$ is the number of nodes in the network. The fragmentation at level~$
\ell $ is $1$ when there is no effective branching and increases when the
split communities at level~$\ell +1$ become more numerous and evenly sized.

As different networks may have different numbers of levels, comparing fragmentation across levels between networks is not straightforward. In addition to the fragmentation scores for consecutive pairs of levels, we propose two measures of global multiscale fragmentation: the \emph{average fragmentation score}
is the mean of the scores for all consecutive pairs of levels, and the \emph{overall fragmentation score} is the branching factor between the coarsest
and most granular levels.

\subsection{Application of EBF to multiscale structural political fragmentation}
\label{sec2.3}

To interpret algorithmic concepts such as scale, resolution, and partition
levels in social terms, we draw on Blau's theory of multidimensional social
structure \cite{blau1977macrosociological,blau2018crosscutting}. Blau
argues that individuals occupy positions simultaneously across multiple
social dimensions, and that patterns of social interaction are shaped by
the degree of similarity and overlap across these dimensions. Individuals
who share more social characteristics are generally more likely to interact,
whereas cross-cutting differences can create ties across otherwise distinct
groups. From this perspective, community structures observed at different
resolutions may reflect different degrees of overlap among individuals'
social and political attributes.

Our previous work provides empirical support for this interpretation by
showing that communities can indeed be organized along multiple social
and political dimensions. However, these dimensions do not map perfectly
onto specific levels of resolution \cite{zhang2025multilevel}. Here, we
provide a brief toy example to illustrate one possible social interpretation
of our measure. While this interpretation is plausible and supported by
social theory and empirical evidence, we recognize that other mechanisms
and external factors may also contribute to the observed multiscale structures.

\begin{figure}[htbp]
\centering
\includegraphics[width=\textwidth]{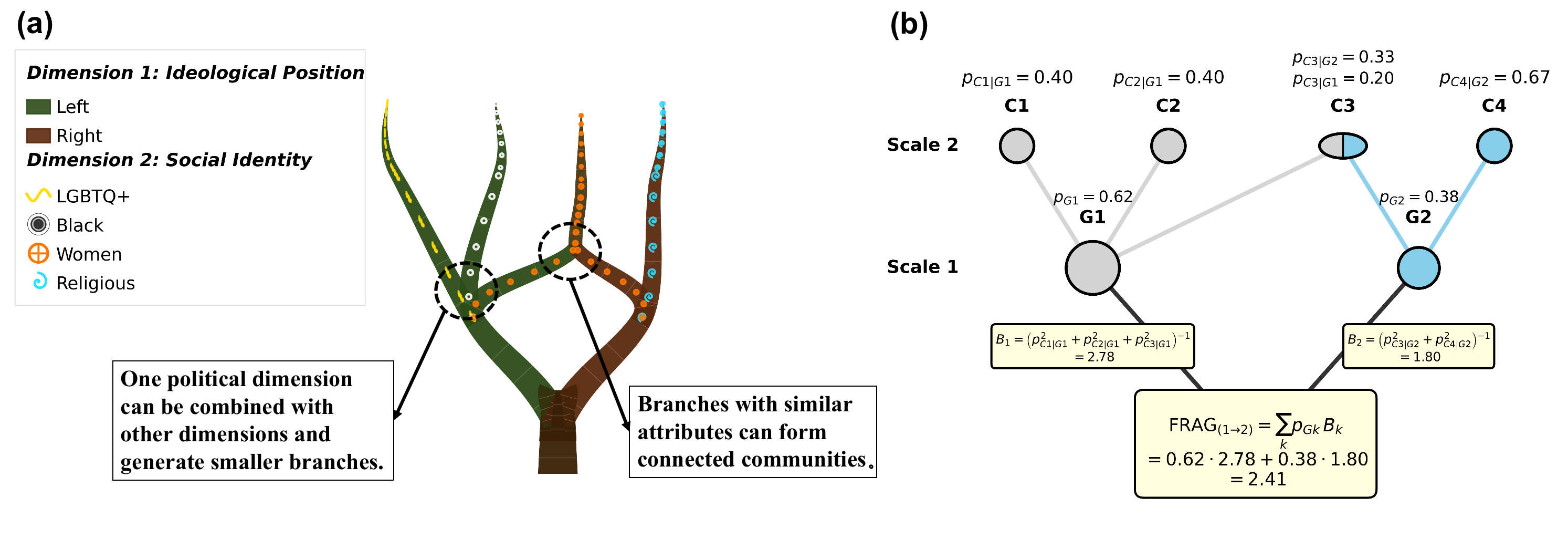}
\caption{Illustration of non-hierarchical
multiscale political fragmentation in online environments and how it can
be measured using the Effective Branching Factor (EBF). Panel (a) illustrates
multiscale structure across ideological and social-identity dimensions.
Panel (b) illustrates the calculation of the EBF for quantifying multiscale
structure}
\label{fig:simulation_multi}
\end{figure}

Suppose two dimensions coexist in the online sphere (e.g., ideological
position and social identity), Fig.~\ref{fig:simulation_multi} (a) shows
a toy example illustrating how communities can be reorganized at different
scales. The categories [left] and [right] on the ideological dimension
can be further combined with the categories [LGBTQ$+$], [Black], [Women],
and [Religious] on the social identity dimension, thereby creating finer
branches at a more granular scale. However, these finer branches may have a non-nested structure. Individuals who share the same social identity, such as [Women], may form a single community at this granular scale and belong to different ideological communities at a coarser scale. 

With our approach (EBF) shown in Fig.~\ref{fig:simulation_multi} (b). We
can calculate the effective sizes of branches at each level that stem from
the previous level, thereby capturing cross-level branching.

In multiscale structures, the effective branching factor provides an additional
measure of fragmentation beyond the effective number of communities and
the within-/between-community edge probability ratio captured at a single
scale.

\section{Application to empirical data}
\label{sec3}

\subsection{Data collection and annotation of empirical datasets}
\label{sec3.1}

We show how our approach can help measure and compare network fragmentation
on two empirical network datasets collected for Brazil, Spain, and the
United States. One dataset comprises online users and the political news
articles they visited, and another dataset is a subset of the online users
and the political influencers they followed on Twitter/X. To collect the
data, we fielded three national surveys conducted during major election
periods: the 2022 Brazilian Presidential Election (2--30 October 2022),
the 2023 Spanish General Election (23 July 2023), and the 2024 U.S. Presidential
Election (5 November 2024). The browsing activity of the survey participants
is tracked during the election with their consent. We then scraped all
public news articles they visited and annotated their left-right ideological
orientation using large language models (LLMs), and validated the annotations
against a media bias-checking platform (see details in Sects.~3 of the Supplementary Information).

Second, Twitter/X handles are voluntarily provided by survey participants
(271 valid handles for Brazil, 217 for Spain, and 279 for the United States).
Based on the donated handles, we collect the political influencers followed
by these users on Twitter/X. Political influencers are annotated by both
human coders and LLMs based on the self-presented information in their
profiles. Human coders first annotate 200 profiles, which are used to validate
the LLM annotations. The LLMs then annotate the remaining profiles. We
also annotate the ideological positions (left, center, or right) of each
political influencer. This procedure identifies 2,307 Brazilian, 4,077
Spanish, and 11,941 U.S. political influencers. The procedures for validating
sample representativeness and identification of political influencers are
detailed in Sects.~4 of the Supplementary Information.

\subsection{Network construction and multiscale community detection}
\label{sec3.2}

We construct a bipartite graph
$ \mathcal{G} = (\mathcal{C}, \mathcal{S}, \mathcal{E}) $ for both datasets,
where $ \mathcal{C} $ denotes ordinary users, $ \mathcal{S} $ political
news articles/influencers, and $ \mathcal{E} $ the visiting/following relationships.
Projecting onto articles/influencers yields a co-follow graph
$ \mathcal{G}^{\mathcal{S}} $, where two articles/influencers are linked
if they have at least one shared visitor or share at least one follower,
with edge weights $ w_{ij} $ indicating the number of shared visitors or shared followers $i$ and $j$.

On this projected network, we apply a multiscale community detection method
to identify clusters of articles co-visited and influencers co-followed
by similar users. Specifically, we use Markov Stability
\cite{delvenne2010stability,lambiotte2015random} with automatic scale
selection \cite{arnaudon2024pygenstability}, optimizing partitions with
the Leiden algorithm \cite{Traag2019}. Communities are defined as regions
retaining random-walk flow over time, with optimal scales determined by
minimizing the Normalized Variation of Information (NVI) across ensembles
of partitions \cite{lambiotte2015random,arnaudon2024pygenstability}. Because
this approach identifies multiple partition levels from smaller to larger
scales, higher levels of partitioning theoretically correspond to greater
resolution granularity and smaller community sizes. This makes fragmentation
across partition levels comparable because we analyze all networks using
the same network construction procedure, Markov Stability approach, and
robustness-based partition-selection criterion. For each network, we determine
the Markov stability scale search range through preliminary exploration
to ensure coverage from highly granular to the coarsest meaningful structural
resolutions. Differences in the number of selected levels therefore reflect
variation in the robust multiscale structures identified within each network
rather than arbitrary differences in the choice of search range. We compare
how community structure changes from finer to coarser patterns within each
country, rather than assuming that a particular level in one country represents
exactly the same type of structure as the same-numbered level in another
country (see more details in Sect.~5 of the Supplementary Information).

\subsection{Empirical case 1: online user - political news consumption network}
\label{sec3.3}

\begin{figure}[htbp]
\centering
\includegraphics[width=\textwidth]{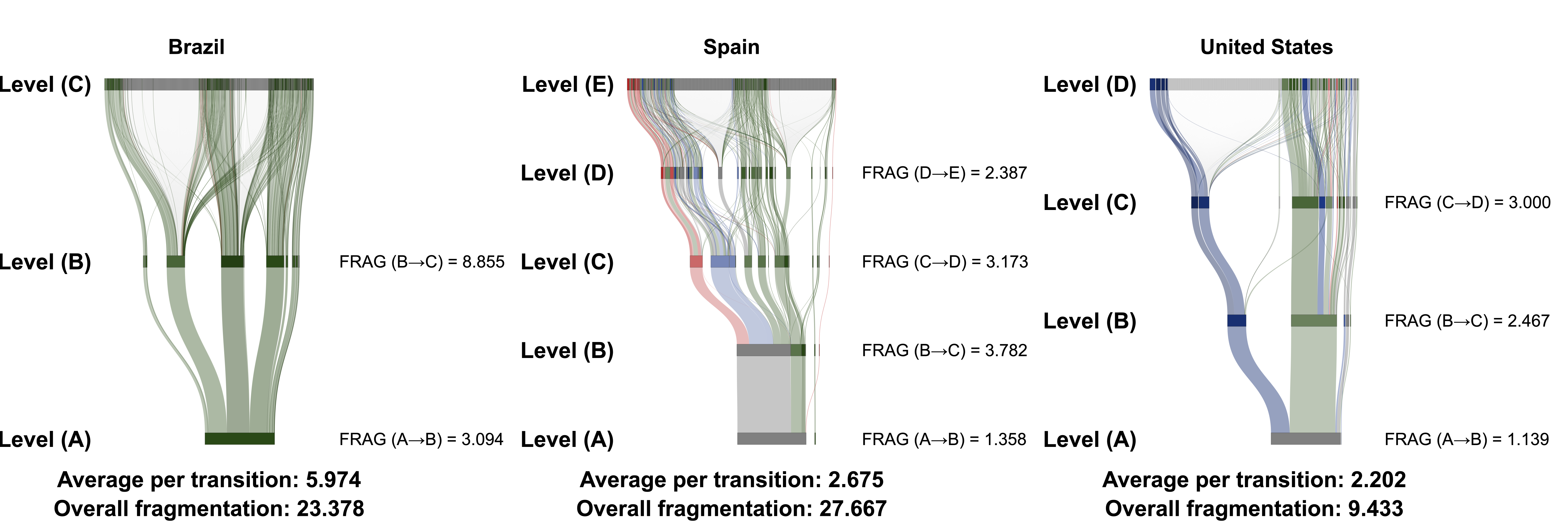}
\caption{Alluvial diagrams of multi-level communities of online political
news in Brazil, Spain, and the United States, with red color indicating
that the majority of the self-reported ideological identities are left-leaning,
blue color indicating that the majority are right-leaning, and green color
indicating that the majority are in a central position. The transparency
of the colors represents the proportions of the majority ideological identities
in each community. The multiscale fragmentation score using EBF for each
level is shown, along with the average and overall fragmentation scores (bottom)}
\label{fig:empirical_case1}
\end{figure}

We apply Markov Stability--based multiscale community detection together
with our EBF measure to online users' political news browsing data from
Brazil, Spain, and the United States to reveal the structural fragmentation
of political news consumption. The multiscale community detection identifies
three levels in the Brazilian network, five levels in the Spanish network,
and four levels in the U.S. network (see Fig.~\ref{fig:empirical_case1}).
Article-level ideological annotations indicate that Brazilian users' news
consumption is concentrated around politically centrist outlets; Spanish
users form several left- and right-leaning clusters, primarily at finer-grained
levels; and U.S. users exhibit clearly defined right-leaning clusters of
news consumption. The average EBF values indicate that Brazil exhibits
the highest fragmentation in political news consumption (6.0). By contrast,
Spain (2.7) and the United States (2.2) display less fragmentation than
Brazil, with Spain being slightly more fragmented than the United States.
However, the overall EBF values indicate that Spain is the most fragmented
(27.7), followed by Brazil (23.4), while the United States (9.4) remains
the least fragmented.

%

The between-level fragmentation scores are annotated at each level in Fig.~\ref{fig:empirical_case1}.
Brazil exhibits a much higher maximum branching value than Spain and the
United States in media consumption, suggesting that a wider variety of
content-category combinations can group Brazilian users together.

\begin{table*}
\caption{Fragmentation by country in online user--political news consumption
networks. For each country and retained level, the table reports the effective
number of communities (ENC), the adaptive E--I index, and the denoised
adaptive E--I index. Levels are ordered from coarsest to finest retained
partition. Values after $\pm $ indicate the standard deviation of the pairwise
adaptive E--I index from the null model}
\label{tab:frag_empirical_1}
\begin{tabular*}{\textwidth}{@{\extracolsep{\fill}}lllll@{}}
\hline
Country & Level & ENC & Adaptive E--I & Denoised Adaptive E--I (Salloum et al.) \\
\hline
 \multirow{3}{*}{Brazil}
& A & 1.191  & $-$0.999 & $-0.551 \pm 0.042$ \\
& B & 3.862  & $-$0.995 & $-0.527 \pm 0.018$ \\
& C & 28.341 & $-$0.989 & $-0.591 \pm 0.013$ \\
[3pt]
\multirow{5}{*}{Spain}
& A & 1.180  & $-$1.000 & $-0.604 \pm 0.077$ \\
& B & 1.637  & $-$0.998 & $-0.781 \pm 0.069$ \\
& C & 6.477  & $-$0.995 & $-0.714 \pm 0.050$ \\
& D & 18.051 & $-$0.994 & $-0.718 \pm 0.022$ \\
& E & 32.988 & $-$0.996 & $-0.671 \pm 0.021$ \\
[3pt]
\multirow{4}{*}{United States}
& A & 1.702  & $-$0.999 & $-0.463 \pm 0.104$ \\
& B & 1.985  & $-$0.998 & $-0.672 \pm 0.011$ \\
& C & 5.105  & $-$0.997 & $-0.590 \pm 0.029$ \\
& D & 16.005 & $-$0.996 & $-0.616 \pm 0.017$ \\
\hline
\end{tabular*}
\end{table*}

For each level, we also report the effective number of fragments, measured
by the ENC, and the strength of separation, measured by the pairwise adaptive
E--I index, as complementary indicators to EBF (see Tab.~\ref{tab:frag_empirical_1}).
We also report the denoised version of the pairwise adaptive E--I index,
as proposed by Salloum et al. \cite{salloum2022separating}. The following
comparison is primarily based on the values from the denoised version.
The results show that, at the finest level, Spain has the largest effective
number of fragments in users' media consumption behavior (33.0), followed
by Brazil (28.3), while the United States has fewer effective fragments
(16.0). It aligns with recent studies showing that the expansion of digital
media has revealed more multidimensional political clusters, such as class,
religion, and center--periphery cleavages
\cite{casan2022online,zhang2025multilevel,ramaciotti2024american}. Spain
and Brazil show more branches than the U.S., possibly because their multi-party
systems provide more fragmented and multidimensional political space than
the two-party system \cite{gross1984comparing}. This is also consistent
with comparative media theory
\cite{hallin2004comparing,humprechtMediaSystemsDigital2022}, which highlights
the higher level of political parallelism, the extent to which media systems
reflect political divisions, in Spain compared to the U.S.

With respect to the largest separation strength observed in each country,
indicated by the most negative values of the denoised pairwise adaptive E--I index,
communities in Brazil have the strongest separation at the finest level
(C). Spain demonstrates the strongest separation at the second level (B),
where the annotations indicate a separation between a large mixed community
comprising the majority of ideologically biased news and other central
news consumption (see Fig.~\ref{fig:empirical_case1}). In the United States,
the strongest separation also appears at the second level (B). The annotations
in Fig.~\ref{fig:empirical_case1} illustrate that the second level in the
United States demonstrates a clear separation between right-leaning and
other news consumption. These cross-national differences may be partially
associated with variations in political and party systems. For instance,
prior work has shown right-leaning asymmetries in political polarization
in the United States \cite{gonzalez2023asymmetric}, whereas Spain exhibits
a more ideologically balanced structure of political competition
\cite{barbera2015birds}. In contrast, Brazil's political landscape is shaped
by a more fragmented party system and salient demographic cleavages
\cite{layton2021demographic}, potentially leading to more complex, multidimensional
patterns of fragmentation.

Comparing the most negative denoised pairwise adaptive E--I values across the three countries,
Spain shows the strongest maximum separation, followed by the United States,
whereas Brazil has the weakest. Taken together, the E--I, ENC, and EBF
measures provide a full picture of the multiscale fragmentation of these
networks and enable their comparison and the identification of their particularities.

In Sect.~6 of the Supplementary Information, we reproduce our results
using different projection methods for bipartite networks: a similarity-normalized
projection \cite{borgatti2011analyzing}, an activity-adjusted projection
\cite{newman2001scientific}, and a binary projection. While the resulting
networks differ, the average EBF results and ranking of fragmentation across
countries remain similar for the similarity-normalized and binary projections.
However, the activity-adjusted projection gives Brazil as less fragmented than the other countries.


\subsection{Empirical case 2: online user - political influencer following network}
\label{sec3.4}

\begin{figure}[htbp]
\centering
\includegraphics[width=\textwidth]{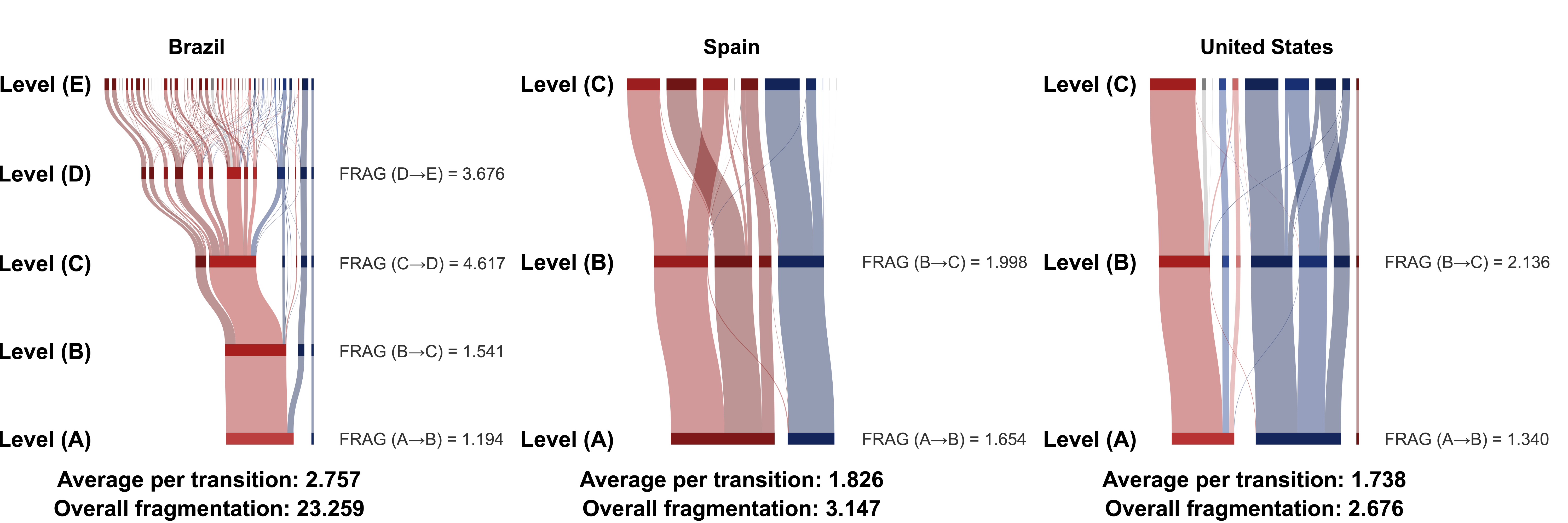}
\caption{Alluvial diagrams of multi-level communities of Twitter/X online
political influencers in Brazil, Spain, and the United States, with red
color indicating that the majority of the self-reported ideological identities
are left-leaning, blue color indicating that the majority are right-leaning,
and green color indicating that the majority are in a central position.
No community has a majority of members self-identifying as centrist. The
transparency of the colors represents the proportions of the majority ideological
identities in each community. The multiscale fragmentation score for each
level is shown, along with the overall fragmentation score (bottom)}
\label{fig:empirical_case2}
\end{figure}

Applying multiscale community detection to the user--political influencer
co-following networks yields seven levels of division in Brazil and four
levels in both Spain and the United States. Fig.~\ref{fig:empirical_case2}
reports the majority ideological orientation (left/center/right) within
each community. Because this network is relatively sparse, we exclude levels
at which more than 90\% of communities consist of a single influencer (see
Fig. S13 in Supplementary Information). After this filtering, five levels
remain for Brazil, and three levels remain for Spain and the United States.

Unlike the political news consumption data, the alluvial diagrams of the
Twitter/X following networks exhibit clearer left- and right-leaning clusters,
with fewer centrist clusters. This pattern is plausible because news articles
are typically framed as relatively neutral reporting, whereas political
influencers more often express explicit ideological positions.

Interestingly, the average EBF scores yield results similar to those from
the political news consumption data: fragmentation is highest in Brazil
(2.8), while Spain (1.8) and the United States (1.7) are less fragmented,
with Spain exhibiting a slightly higher score. However, the overall EBF
scores reveal a different pattern from that observed in political news
consumption networks, identifying Brazil (23.3) as the most fragmented
when fragmentation is assessed from the coarsest to the finest scale. The
level-wise transitional FRAG scores show that Brazil still exhibits the
greatest maximum branching among the three countries (see Fig.~\ref{fig:empirical_case2}).

The single-scale measures, ENC, adaptive E-I index, and denoised adaptive
E-I index, are shown in Tab.~\ref{tab:frag_empirical_2}. Unlike the results
for online users' political news consumption, the Brazilian influencer-following network has the
largest number of fragments at the finest level (24.9), which is almost
five times as large as in Spain (5.4) and the United States (5.2).

Regarding the strength of separation for each country, the denoised adaptive
E--I index indicates the strongest separation at the finest level (E) ($-$0.85)
in Brazil, the first level (A) ($-$0.95) in Spain, and the second level (B)
($-$0.67) in the United States. Fig.~\ref{fig:empirical_case2} shows that
Spain exhibits the strongest separation when there is a clear division
between the left and right, which is consistent with previous studies identifying
Spain as a ``polarized pluralist'' system
\cite{hallin2004comparing,humprechtMediaSystemsDigital2022}. However, the
Twitter/X following network of users in the United States shows the strongest
separation at the middle level, suggesting greater internal separation
within ideological groups in the U.S. Twitter/X space (especially among
right-leaning users) than in Brazil and Spain.

Regarding the comparison of the most negative denoised adaptive E--I values
across countries, Spain still ranks as having the strongest maximum separation
strength. However, Brazil exhibits a stronger maximum separation strength
than the United States, compared to media consumption.

The results are generally robust to variations in the bipartite projection
method (see Sect.~6 of the Supplementary Information). The ranking
of the average EBF across countries is the same for all projection methods.
Only the overall EBF yields a different ranking with the activity-adjusted
projection, making the US network appear more fragmented than the Spanish
one.


\begin{table*}
\caption{Fragmentation by country in online user--political influencer following
networks. For each country and retained level, the table reports the effective
number of communities (ENC), the pairwise adaptive E--I index, and the
denoised adaptive E--I index. Levels are ordered from coarsest to finest
retained partition. Values after $\pm $ indicate the standard deviation}
\label{tab:frag_empirical_2}
\begin{tabular*}{\textwidth}{@{\extracolsep{\fill}}lllll@{}}
\hline
Country & Level & ENC & Adaptive E--I & Denoised Adaptive E--I (Salloum et al.) \\
\hline
\multirow{5}{*}{Brazil}
& A & 1.064  & $-$1.000 & $-0.800 \pm 0.015$ \\
& B & 1.276  & $-$0.994 & $-0.791 \pm 0.010$ \\
& C & 2.047  & $-$0.994 & $-0.573 \pm 0.048$ \\
& D & 9.685  & $-$0.959 & $-0.771 \pm 0.014$ \\
& E & 24.912 & $-$0.747 & $-0.850 \pm 0.024$ \\
[3pt]
\multirow{3}{*}{Spain}
& A & 1.750 & $-$0.957 & $-0.951 \pm 0.001$ \\
& B & 3.418 & $-$0.903 & $-0.828 \pm 0.001$ \\
& C & 5.372 & $-$0.414 & $-0.706 \pm 0.031$ \\
[3pt]
\multirow{3}{*}{United States}
& A & 2.019 & $-$0.988 & $-0.561 \pm 0.006$ \\
& B & 4.155 & $-$0.967 & $-0.669 \pm 0.004$ \\
& C & 5.169 & $-$0.967 & $-0.476 \pm 0.009$ \\
\hline
\end{tabular*}
\end{table*}

To better contextualize the different EBFs we obtain from online activity,
we also compute the effective number of parties using ENC based on the distribution
of seats in the lower chambers of Brazil, Spain, and the United States
following the 2022, 2023, and 2024 elections, obtaining values of 11.8,
3.4, and 2.0, respectively. Notably, the cross-national ordering of average multiscale
fragmentation online aligns with these electoral outcomes. However, Brazilian
social media users appear less fragmented than the party-seat distribution
alone would suggest, likely because contestation between Bolsonaro and
Lula makes bipolarization more visible online.

\section{Conclusion and discussion}
\label{sec4}

As complex real-world social and political systems evolve, there is a growing
need for network measures that accurately capture both single-scale and
multiscale network fragmentation. To meet this need, this study first used
analytical analyses and simulations to assess how key parameters shape
six existing fragmentation measures in the single-scale setting: modularity,
the global-level, community-level, pairwise, and pairwise adaptive E--I
indices, and ENC.

Our findings show that the pairwise adaptive E--I index is the most appropriate
measure of community separation strength, as it is independent of the network
size and the number of communities and accounts for communities of different
sizes. Complementing this measure by the effective number of communities,
ENC, provides a complete description of the network fragmentation in the
single-scale setting. To capture multiscale fragmentation, we introduced
the effective branching factor, EBF, which captures both intra-level diversity
and inter-level branching.

Empirical applications of the average EBF to two cross-national datasets
of online behavior produced consistent results across datasets and were
in line with the fragmentation measured at the level of party seats in
the lower chambers. Brazil was the most fragmented in terms of EBF, Spain
showed an intermediate level, and the United States was the least fragmented.
However, Brazil appears to be less fragmented online than electoral voting
behavior would suggest.

Although the multiscale comparison reveals similarities between the two
datasets, the single-level-wise ENC and pairwise adaptive E-I index provide
additional insight into their differences. For example, Spanish political
news consumption exhibits a substantially larger effective number of fragments
than Spanish users' political influencer following on Twitter/X, relative
to Brazil and the United States. Moreover, Brazil does not show a strong
separation between ideological groups in online political news consumption,
and Spain shows only a slight separation at finer levels, whereas such
separation is more pronounced in their Twitter/X political influencer-following
networks. In the United States, users exhibit a pronounced separation pattern
between the right-leaning and other clusters in political news consumption,
whereas the strongest separation in the Twitter/X political influencer-following
network is characterized by both left- and right-leaning clusters.

These results should be interpreted in light of several limitations. First,
because no other network-based metrics have been developed to quantify
fragmentation in multiscale settings, we contextualize our approach using
(1) applications to two different empirical datasets, (2) a single-scale
measure based on seat shares, and (3) visualizations. Although the average EBF score rankings are consistent across datasets and approaches,
we cannot externally validate finer-grained outputs, such as level-specific
branching capacity, because these details are not available in any previous
methods. Moreover, our examples rely on bipartite projections, which can
yield different results depending on the projection method used. We find
that the ranking of fragmentation was generally robust across projection
methods, but changed when we discounted users' activity in the projection.
While this does not invalidate our fragmentation measures, it highlights
the importance of the projection method and its impact on the measured
fragmentation. Second, ideological annotations in the two empirical datasets
rely primarily on LLM-based classification. While we validated the U.S.
outlet-level labels in the political news consumption dataset using ratings
from media-bias assessment platforms, we lack comparable validation for
the other countries and for the political influencer co-following dataset.
Finally, we analyze only two empirical datasets, political news browsing
behavior and Twitter/X political influencer following behavior, which limits
the generalizability of the results. Future work should apply the proposed
approach to additional forms of structural behavior.

Despite these limitations, this study contributes to measuring political
fragmentation in both single-scale and multiscale spaces. Network-based
approaches offer significant potential for investigating complex real-world
patterns. However, inappropriate methodological choices can yield biased
estimates and misleading conclusions, and may even have adverse societal
consequences when used to inform policy recommendations. For example, if
a measure confounds true political separation with group-size imbalance
or the number of detected communities, or if a single-scale measure is
used to assess a multiscale political structure, it may overestimate or
underestimate fragmentation, thereby distorting the apparent degree of
division of a political system. It is therefore essential to understand
which aspects these measures capture and which they do not in real-world
settings. This work also shows that none of the measures capture all aspects
of network fragmentation; therefore, scholars should report all three aspects
in future studies of political fragmentation phenomena.

Finally, our proposed metric takes only a multilevel partitioning of a
network (hierarchical or non-hierarchical) as input and is therefore potentially
applicable to any network where such information is available, for example,
when produced by a multiscale community detection method.

\section*{Acknowledgements}
Y.Z. and A.B. thank Dorian Quelle, Yasaman Asgari, and Samuel Koovely for the interesting discussions and their valuable feedback.

\section*{Author contributions}
Y.Z. and A.B. designed the research and the methodology. Y.Z. performed the data analysis and wrote the original draft; A.B. supervised the research; L.C.H. and F.E. supervised the political and communication theory framework; A.B., L.C.H., and F.E. reviewed and edited the draft.

\section*{Funding}
This work is supported in part by funds from the Swiss National Science Foundation grant 100017\_204483.

\section*{Data availability}
The data and code underlying this article are available in Harvard Dataverse at \url{https://doi.org/10.7910/DVN/CUSTYX}. Supplementary material is available online.

\section*{Ethics approval and consent to participate}
This study analyzes platform data from survey participants who were tracked voluntarily and donated their Twitter/X handles with informed consent and fair compensation; only public profile metadata and follower relations were collected.

This study has received institutional ethics approval (IRB/ethics approval: Faculty of Arts and Social Sciences Ethics Committee, University of Zurich). Data were pseudonymized with restricted access to identifiers, and we report only aggregates.

\section*{Competing interests}
The authors declare no competing interests.

\bibliographystyle{plain}
\bibliography{reference}

\end{document}